# RFID Authentication Protocol Based on a Novel EPC Gen2 PRNG

Pino Caballero-Gil, Cándido Caballero-Gil and Jezabel Molina-Gil

*Department of Statistics, Operations Research and Computing*
*University of La Laguna, Spain*
*E-mail: {pcaballe, ccabgil, jmmolina}@ull.es*

**Abstract**

Continuous advances of Information Technologies (ITs), and in particular of the RFID technologies that allow the connection between the physical world objects and the IT infrastructure, have guaranteed the improvement and efficiency of industrial technologies in the last decades. This paper includes the proposal of two new schemes for RFID. On the one hand, it describes the internals of a lightweight Pseudo-Random Number Generator (PRNG) suitable for low resource devices such as passive RFID complying with the EPC Gen2 specifications. On the other hand, a new secure mutual authentication protocol for such RFID devices that uses the proposed PRNG is presented. The design of the proposed PRNG is based on a nonlinear filter of a Linear Feedback Shift Register (LFSR), and the authentication protocol is lightweight. Both schemes fulfill all practical requirements of low-cost RFID such as resource limitation of EPC Gen2 tags. This is thanks to that only simple computation modules such as the proposed LFSR-based pseudorandom generator and bitwise operations are required. The combination of both proposals guarantees at the same time low power consumption and secure features such as authentication, confidentiality and anonymity.

## 1. Introduction

RFID (Radio Frequency IDentification) is a type of technology used to identify physical objects through radio communications. The increasing applications of RFID cover all aspects of human lives from daily activities to many different industrial applications involving humans, animals and products. This technology provides many opportunities and advantages for users, so it is gradually replacing other systems such as barcodes. One of its main uses is in intelligent manufacturing, where the collection and synchronization of manufacturing data is done through the combination of RFID devices with wireless networks.

There are three main components in any RFID system: tags, readers and a back-end server. Each tag is typically made up of an antenna for receiving and transmitting a radio-frequency signal, an integrated circuit for modulating and demodulating the signal, and limited computational and storage capabilities. Each reader queries tags to communicate with them through wireless communications, and sends the received information to the back-end server through a secure channel. The server is composed of some processors and a database containing information about the tags it manages. Tags and readers are connected through radio communication whilst readers and server are connected through a secure channel.

RFID systems can be classified depending on whether the tags are active or passive. Active tags have their own battery power source, while passive tags have no battery and are activated by the electromagnetic waves of the reader. Passive tags are smaller, lighter and cheaper than active tags. Both have limited memory capacity and computing power, but in particular passive tags have strict limitations in their resources. The RFID standard EPC Class 1 Generation 2 (EPC Gen2) was approved in 2004 to address several problems with tags of previous RFID protocols [11]. According to it, tags are passive, so they are considered unsafe and susceptible to physical attacks. This means that security should be considered in the design of any EPC Gen2 compliant protocol in order to use it securely in its many applications. Indeed, security is perceived as an influential aspect on customer trust [24] [28].

Compared to traditional wireless networks [15], low-cost RFID tags nature, i.e. its restricted computing capacity and its limited memory capacity make RFID systems more vulnerable to attacks. Consequently, several security problems in RFID systems must be solved before using such technology securely. First, the unauthorized access to any tag ID information must be avoided. Secondly, it must be prevented that potential adversaries can mislead easily the reader by using gathered ID information of valid tags. These two security problems can be respectively solved through encryption and authentication. In order to achieve confidentiality, the creation of a secure communication channel in an insecure environment is required. Symmetric encryption is proposed here as the most efficient method to establish a secure communication channel between tag and reader. In general, logic gates and memory of passive tags are not sufficient to use standardized cryptographic algorithms such as AES [12]. Indeed, according to the EPC Gen2 standard, tags only support on-chip a 16-bit Pseudo-Random Number Generator (PRNG). Thus, a new lightweight nonlinear PRNG complying such a condition is here proposed to compute a secret key for a stream cipher. On the other hand, a novel interactive mutual authentication scheme based on that PRNG is proposed for its use in low-cost EPC Gen2 compliant RFID communications.

The rest of this paper is organized as follows. Next section includes a short state of the art where some weaknesses of existing schemes are commented. Section 3 introduces some details of the EPC Gen2 standard. The main characteristics of our proposals are described in section 4. Section 5 provides the detailed description of the novel pseudo-random number generator conforming to the EPC Gen2 standard and a full study about the produced sequences. Section 6 includes a new mutual authentication scheme based on the proposed generator, and designed to address privacy and security requirements of low-cost RFID and its analysis. The last section concludes the paper and points out some future works.

## 2. Related Work

In most RFID works the approach is quite practical because the number of existent applications is enormous [22]. A comprehensive repository of published papers on the topic can be reached online at [1]. For instance, [27] included a scheme based on a hash function and a pseudorandom number generator in order to prevent tracking. Despite the fact that hash functions can be implemented efficiently on low-power hardware, they are still far from the current capacities of low-cost RFID tags, as indicated in [17]. Indeed, just a few security schemes presented in the literature meet the EPC Gen2 standard. For instance, some proposals for authentication between reader and tag use pseudo-random functions or PRNGs [3], but none of them are really optimized for EPC Gen2 compliance. The protocol proposed in [9] uses two types of keys to defend against DoS (Denegation of Service) attacks that cause interruptions of synchronization between the back-end server and the tags. Furthermore, the scheme is vulnerable to information leakage and replay attacks. [8] included a lightweight mutual authentication protocol for solving the secret disclosure problem and the replay attack in the scheme proposed in [21]. Nevertheless, the cloning attack problem is not yet solved for this scheme. [4] proposed a lightweight mutual authentication protocol compatible with session unlinkability, forward and backward secrecy. Their proposal is optimistic with constant key-lookup, and can be easily implemented on EPCGen2. Still, it is susceptible to replay and cloning attacks. [7] included in a mutual authentication protocol that reduces the workload on the database and ensures privacy. However, they did not consider cloning attacks. Many anti-counterfeiting schemes have been recently proposed because cloning attack issues are common in low-cost RFID tags. In [10] a serial number was used for all tags to prevent cloning attacks. Nevertheless, the scheme does not conform exactly to EPC Gen2 standard because it uses a 32-bit PRNG instead of a 16-bit PRNG. [16] proposed several simple authentication techniques for preventing skimming attacks, but it did not consider eavesdropping and privacy invasion threats.

With respect to pseudorandom number generators for EPCGen2, [23] proposed LAMED, an algorithm initially based on 32-bit keys. In [14] it was showed a method to derive random data from the initial state of tag memory. Also [2] proposed the extraction of randomness from sampling radio signals. [6] described a hybrid method to build random sequences by combining physical properties and the use of linear feedback shift registers [5]. It is worth mentioning that our proposal meets all EPC Gen2 requirements, its security is guaranteed, and in general it has a lower hardware complexity than the aforementioned proposals.

Many researchers have studied mutual authentication based on the server. Various protocols are based on different tools such as hash functions, message authentication codes, block ciphers, pseudo-random functions, etc. [20] proposed the use of synchronized secrets so that the back-end server gives through the reader to the tag a new random number every time the tag is used. However, the scheme seems susceptible to eavesdropping and impersonation.

An analysis of server-less RFID authentication protocol that allows readers to authenticate tags without the help of any online back-end server was presented in [19]. [25] proposed a scheme based on a challenge-response approach and simple shifts and bit-wise operations together with a keyed hash function. Such a scheme is vulnerable to tag impersonation and server impersonation attacks. [26] proposed an authentication protocol based on a challenge-response scheme and secret key cryptography where a non-volatile state is required on the tag, which might be too expensive for a low-cost tag. In this paper we present new solutions for pseudorandom generation and mutual authentication conforming to the EPC Gen2 Standard, which are completely different from all the aforementioned proposals.

Therefore, both the generator and the authentication design in EPC Gen2 RFID can be considered problems still open because most proposals require too many resources and/or do not fulfill the standard and/or are not secure against different attacks.

## 3. The EPC Gen2 Standard

The Electronic Product Code (EPC) is a fully established standard in RFID area, where it has also emerged the EPCglobal Network initiative to connect software and real objects. In 2004, EPCglobal ratified the standard EPC Class 1 Gen 2 (EPC Gen2) for RFID implementations. According to this standard, tags are passive, so they have their computational capabilities very restricted. Tags communicate with RFID readers in UHF band and its communication range can be from 2 to 10m. Also tag memory is limited and must be considered unsafe and susceptible to physical attacks. In that standard, tags only support on-chip a 16-bit Pseudo-Random Number Generator (PRNG) and a 16-bit Cyclic Redundancy Code (CRC). There is a 32-bit kill command used to disable the tag permanently, and a 32-bit access PIN to access its internal memory. Despite the large progress that the EPC Gen2 standard implies in terms of communication compatibility and performance, and its significant impact for the dissemination of RFID technology, its security level is very weak.

Two important operations for tag management are included in the EPC Gen2 standard: inventory and access, and both present serious security flaws. In particular, as shown in Figure 1, the inventory protocol is an interactive algorithm between reader and tag with at least 4 steps that include: a Query $\in$ [0, 15], a 16-bit Random Number RN16, an ACKnowledgment ACK(RN16), and tag identifying data EPCdata. If the tag does not receive a valid ACK(RN16), it goes to its initial state and the whole process is repeated.

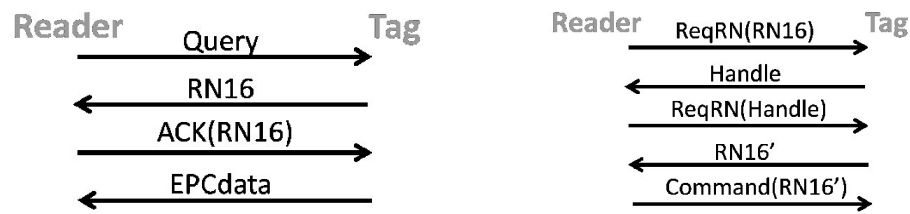

Fig. 1 EPC Gen2 Inventory Protocol and EPC Gen2 Access Protocol

After the inventory protocol, the reader may choose to access thee tag. The access command shown in Figure 1 is also an interactive algorithm between reader and tag with at least 5 steps that include: ReqRN(RN16) containing the previous RN16, a new RN16 denoted Handle, ReqRN(Handle) containing the previous Handle, to which the tag responds with a new RN16'. Finally, the reader then generates a 16-bit ciphertext string with the XOR between the 16-bit word to be transmitted and this new RN16'.

Spoofing and tracking attacks and access to private stored information are possible in the inventory command by simply listening to the radio channel, because the fixed EPCdata are transmitted as clear text. In the command access, security is also extremely weak because passive attacks might be carried out by listening to backward and forward channels, and by taking the random numbers sent by the tag to decrypt the ciphertext sent by the reader.

The so-called Cyclic Redundancy Code CRC in EPC Gen2 standard is a checksum algorithm to detect transmission errors efficiently. It takes as input a binary stream encoded as coefficients of a polynomial that is divided by another fixed polynomial in order to get a rest, which in binary expression, is its output. EPC Gen2 specification proposes the use of $x^{16}+ x^{12}+ x^{5}+ 1$ as fixed polynomial, what allows the detection of all single and double errors, as well as all errors with an odd number of bits, all burst errors of 16 or less contiguous bits, 99,997% of 17-bit error bursts, and 99,998% of 18-bit and longer bursts. The hardware requirements of the CRC algorithm are not large because it is really a shift register of 16 bits. Note that the CRC algorithm is linear, so it cannot be used in cryptographic applications because it does not detect malicious changes.

The pseudo-random number generator in the EPC Gen2 standard is a deterministic function that generates a pseudorandom binary sequence, and consists of two components: a state and a generation algorithm. In order to generate a random bit, the PRNG algorithm takes as input the previous state so that the output is a new state together with the output bit. The EPC Gen2 standard does not detail how to implement this PRNG but describes minimum randomness criteria for the sequences it produces in order to ensure an acceptable pseudorandomness level. When designing a PRNG that fits the EPC Gen2, the strict limitations in hardware have to be taken into account. Note that low-cost RFID tags have approximately 5K-10K gates, and within this number only from 400 to 4K gates can be used for security tasks. In addition, there are strict time requirements as there should be possible to read a number of tags in a given period of time, approximately at a rate of 450 tags/sec. This implies a serious limitation on the maximum number of iterations that a tag can do to generate a pseudorandom number.

Despite security concerns of the EPC Gen2 standard it can be considered a success, according to its adoption by most RFID manufacturers. In this paper, we propose new methods for authentication, encryption and privacy protection in compliance with the standard features described in EPC Gen2. Our schemes do not rely on RFID readers due to their portability. Instead of that, our proposal bases its security on the trust in the back-end server because all shared secrets (like EPCdata) are stored only by tag and back-end server's database, with no possible access by the reader at any time.

## 4. Preliminaries

The first authentication proposals for RFID were based on unilateral authentication. However, mutual authentication is necessary between readers and tags in most applications. Since standard cryptographic protocols such as hash functions, message authentication codes, block ciphers, etc. require too many resources to be used in low-cost RFID tags and surpass their capabilities, it is necessary to define new lightweight cryptographic protocols for the definition of mutual authentication schemes for low-cost RFID.

A typical mutual authentication solution based on a secret key shared between two entities consists in that each entity has to convince the other that it knows the shared secret key. Thus, to prevent tag cloning, a challenge-response scheme based on symmetric key cryptography can be used. This is the main idea behind the mutual authentication scheme here proposed.

Replay attacks represent a possible weakness of RFID technology. In order to prevent them, typical cryptographic solutions are incremental sequence numbers, clock synchronization or nonces. Passive RFID tags cannot use clocks because they do not have any power supply so clock synchronization is not feasible. On the other hand, incremental sequences are not adequate to avoid tracking. Therefore, in the scheme described in this paper we use nonces.

In order to protect data transmitted between tag and reader against eavesdropping, the typical solution is encryption. In particular, the simplest encryption function is the XOR operation used in stream cipher. However, in that case the problem is not encryption, but key generation and management because it is necessary to produce a new encryption key for each session. This is solved with our authentication proposal.

Lastly, to prevent tag tracking, as aforementioned, the update of tag ID can be used. If tag

ID knowledge is only shared between the back-end server and the tag, an easy way to update it is to use the same pseudorandom number generator PRNG both by the tag and by the back-end server (see Figure 2), what implies the need for synchronization between tag and server. Such a tag ID update is used in the scheme here proposed.

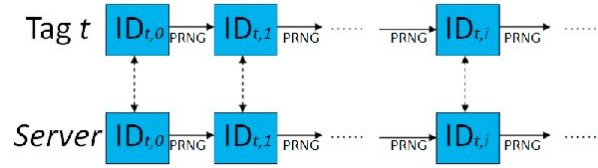

Fig. 2 Tag ID Update

## 5. Description of the Nonlinear PRNG

This section includes the definition of a new PRNG that is to be used in RFID technology both for symmetric encryption through stream cipher and for challenge-response authentication based on symmetric cryptography.

A PRNG can be seen as a deterministic function whose output is computed from the previous output. It is initialized with a randomly chosen seed. The PRNG strength depends on the period and the probability distribution of the output sequences. Many PRNGs are based on LFSRs because they can be applied efficiently in low-cost RFID tags. Here we define a new LFSR-based PRNG that will be used with session keys shared between tags and server.

According to the EPC Gen2 specification, a low-cost RFID tag must be capable of generating 16-bit pseudorandom numbers. In this way, the EPC Gen2 standard is based on a 16-bit PRNG, and the success probability of any attack to any PRNG-based protocol is lowerbounded by $2^{-16}$. Instead, a future version of the EPC Gen2 standard might be based on a 32-bit PRNG in order to improve its security and also to take advantage of the 32-bit PIN defined in the specification of the original EPC Gen2 standard. On the contrary, now the XOR of two halves of the 32-bit PIN provides no better security than a 16-bit PIN.

An LFSR is defined by a feedback function and by a string of binary cells sharing the same clock signal. To produce a bit, the register contents are shifted one position, extracting as output the most significant bit of the register in the previous state. The feedback function allows computing every new bit from some bits of the register, so that this new bit is the new least significant bit of the new state. The feedback function of an LFSR is basically an XOR operation on some contents of the cells of the state, given by the feedback polynomial over GF(2) denoted $C(x)=1+ c_1x+ c_2x^2+ \cdots + c_Lx^L$, whose degree L is the length of the register. The period of sequences produced with an LFSR of length L is less than or equal to $2^L-1$, value that is reached when its feedback polynomial $C(x)$ is primitive, in which case the produced sequences have optimum statistical properties. The LFSR is commonly used as pseudorandom generator in cryptography due to several fine characteristics. First, the characteristics of the produced sequences are optimal. Also its hardware implementation is efficient and its computational requirements are simple. However, LFSRs have significant drawbacks that must be solved in order to be used safely in cryptography. The biggest problem comes from its linearity because the initial state or seed of the LFSR can be easily determined with a simple system of linear equations by using the polynomial function C(x) and a 2L-bit output keystream. Thus, in order to use an LFSR to build a PRNG, the linearity problem should be solved, what in cryptography is usually performed with methods such as the so-called non-linear filtering or the non-linear combination of several LFSRs.

Our proposal is based on an LFSR mainly because it is an ideal system for both energy [18] and computational constrained environments. In addition, note that the CRC described in the EPC Gen2 standard requires the same hardware than the LFSR scheme so the LFSR used in our scheme for the PRNG might be also used for the CRC.

The design is based on an LFSR with primitive feedback polynomial to achieve maximum period and to fulfill the pseudorandomness of the generated sequences. In particular, in the proposal the number of non zero coefficients of the feedback polynomial C(x) is the smallest possible integer greater than 0.2L to avoid correlation attacks and to ensure efficiency.

The proposed generator consists of two main building blocks: an LFSR and a filter function. The order of the nonlinear filter function has been chosen to be the greatest prime number p less than or equal to the number L/2, in order to ensure a large linear complexity.

The nonlinear filter function includes a linear term corresponding to the stage indicated by the function order. The number of terms in each order i= 2, 3, ..., p is given by the integer part of L/i. These terms are obtained by multiplying successive disjoint stages, in order to achieve pseudorandomness and confusion.

Below we specify the concrete details of our design generally sketched in Figure 3. The LFSR used in the design is of length L=16 in order to comply with the EPC Gen2 standard, and its contents are denoted by $s_j$, $s_{j+1}$, ..., $s_{j+15}$. In particular, the proposed feedback polynomial C(x) of the LFSR is a primitive polynomial of degree 16 defined as $C(x)= 1+ x^2+ x^7+ x^9+ x^{16}$ so that the update function is $s_{j+16}= s_{j+14}+ s_{j+9}+ s_{j+7}+ s_j$. The content of the 16-bit LFSR represents the state of the cipher and input of the nonlinear filter function f. In particular, from such a state, 16 variables are taken as input to the Boolean function f(x) of algebraic degree 7. Such a filter function has been chosen to be balanced, first-order correlation-immune, and with high nonlinearity.

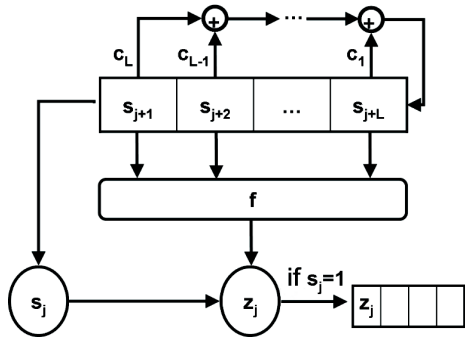

**Fig. 3 General Description of the PRNG**

The filter function is defined as

$f(x_0, x_1, ..., x_{15}) =$

$= x_6 + \sum_{i=0}^{7} x_{2i} x_{2i+1} + \sum_{i=0}^{4} x_{3i} x_{3i+1} x_{3i+2} +$

$+ \sum_{i=0}^{3} x_{4i} x_{4i+1} x_{4i+2} x_{4i+3} + \sum_{i=0}^{2} x_{5i} x_{5i+1} x_{5i+2} x_{5i+3} x_{5i+4} + + \sum_{i=0}^{1} x_{6i} x_{6i+1} x_{6i+2} x_{6i+3} x_{6i+4} x_{6i+5} +$

$+ \sum_{i=0}^{1} x_{7i} x_{7i+1} x_{7i+2} x_{7i+3} x_{7i+4} x_{7i+5} x_{7i+6} = x_6 + x_0 x_1 +$

$+ x_2 x_3 + x_4 x_5 + x_6 x_7 + x_8 x_9 + x_{10} x_{11} + x_{12} x_{13} +$

$+ x_{14} x_{15} + x_0 x_1 x_2 + x_3 x_4 x_5 + x_6 x_7 x_8 + x_9 x_{10} x_{11} +$

$+ x_{12} x_{13} x_{14} + x_0 x_1 x_2 x_3 + x_4 x_5 x_6 x_7 + x_8 x_9 x_{10} x_{11} +$

$+ x_{12} x_{13} x_{14} x_{15} + x_0 x_1 x_2 x_3 x_4 + x_5 x_6 x_7 x_8 x_9 +$

$+ x_{10} x_{11} x_{12} x_{13} x_{14} + x_0 x_1 x_2 x_3 x_4 x_5 + x_6 x_7 x_8 x_9 x_{10} x_{11} + + x_0 x_1 x_2 x_3 x_4 x_5 x_6 + x_7 x_8 x_9 x_{10} x_{11} x_{12} x_{13}$

where the variables $x_0, x_1, ..., x_{15}$ correspond to the tap positions $s_j, s_{j+1}, ..., s_{j+15}$, respectively. The design of the cipher, shown in Figure 4, has been chosen to be as simple as possible for a hardware implementation. Since the LFSR in the cipher is of length 16 and its polynomial is primitive we know that the period of the LFSR keystream is $2^{16} - 1$.

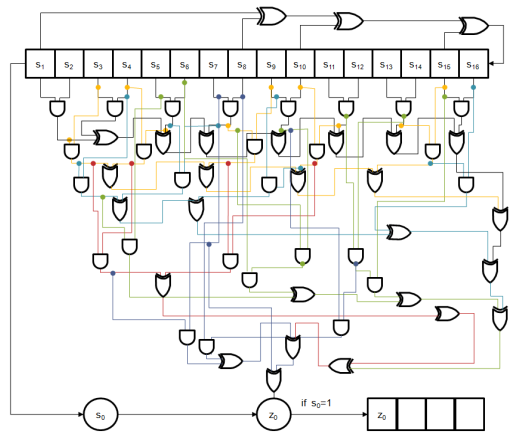

**Fig. 4 Detailed Description of the PRNG**

To avoid correlation attacks, the output of the nonlinear filter is irregularly decimated so that the output of the LFSR determines whether the corresponding output of the filter function is used or discarded. Thus, since the LFSR is regularly clocked, on average the cipher will output 0.5 bit/clock. Finally, to ensure a stable output, a buffer of size 4 is included.

The generator has been designed to be very small in hardware, using as few gates as possible while maintaining high security. Thus, the cipher is intended to be used in environments where gate count, power consumption and memory needs to be very small. This is exactly the case of EPC Gen2 RFID technology. Golomb's randomness postulates [13] are conditions that a sequence should fulfill in order to appear random. A binary sequence that satisfies Golomb's postulates is called a Pseudo-Noise (PN) sequence. Each postulate has an immediate translation into some test of randomness. In order to prove pseudorandomness of our generator, we have generated with it a large number of sequences and submit them to a battery of statistical tests. Since most sequences pass most tests, the confidence in the pseudorandomness of the sequences is large and so is the confidence in the generator. In particular, the generator has passed the following tests: Frequency Test, Serial Test, Poker Test, Runs Test and Autocorrelation Test.

The proposed PRNG has been implemented in software to check the pseudorandomness of the output sequences. We generated 3.9 Gb of data with our PRNG in order to check the three statistical properties described in the EPC Gen2 specification:

1. The probability that any 16-bit sequence produced by the PRNG has value j, for any j, must belong to the interval $[0.8/2^{16}, 1.25/2^{16}]$.
2. The probability that any of two or more tags simultaneously generate the same binary sequence must be less than 0.1%.
3. A given sequence produced by the PRNG 10ms after the end of the transmission must not be predictable with a probability greater than 0.025% if the outcomes of prior outputs from the PRNG are known.

Regarding the first property, our experiments involved $2^{16}-1$ bits produced with our generator for every possible seed of the LFSR. The results of all aforementioned statistical tests for a significance $\alpha = 0.05$ led to the conclusion that our proposed PRNG fulfills the first specification of the EPC Gen2 standard. In particular, the frequency test, which is based on the proportion of zeroes and ones, checks the closeness of the proportion of ones to 0.5. In this case we obtained 86,55% of positive results over all possible inputs. On the other hand, the serial test, whose focus is to determine whether the frequency of all possible $2^m$ m-bit overlapping patterns across the sequence is approximately the same. 100% of all possible outputs pass the serial test. For the poker test we divide the sequence into subsequences of a certain length, and then check whether these sequences appear the same number of times. It results in 80,89% of positive results with our generator over all possible inputs. The purpose of the runs test is to determine whether the number of runs of ones and zeros of various lengths is as expected for a random sequence. In particular, this test produced 97,74% of positive results for short runs, and 100% of positive results for long runs, for the sequences produced with the proposed generator over all possible inputs. Finally, in the autocorrelation test, the purpose is to check for correlations between the sequence and its shifted versions. It produces 100% positive results with the proposed generator over all possible inputs.

To test the second property, ten thousand random and different seeds have been used to initialize the PRNG and the correlation between the obtained sequences has been computed, concluding that two simultaneous identical sequences do not appear.

The third property is related to the probability of prediction. In order to check that this property holds, the serial correlation test has been implemented. Such a test measures the extent to which each m-bit output depends upon the previous m-bit output. For our sequences, this value is obtained close to zero so we conclude the fulfillment of the property.

Furthermore, we computed the linear complexity of the produced sequences through the Berlekamp-Massey Algorithm over $2^{16}$-1 bits of the keystream, confirming that it is always maximum and equal to half the number of analyzed bits. Also the period was computed obtaining always values around $2^{16}$-1 for every possible seed.

Several general attacks on stream ciphers are now discussed in order to investigate to what extent they can be applied against the proposed generator. Indeed, resistance against known cryptanalytic attacks is the most important basis for the design of a new encryption algorithm because there should be no faster successful attack than the exhaustive key search.

Due to the good statistical properties of the PN-sequences produced by the LFSR, and because the function f is first-order correlation-immune, the correlations between the outputs of the generator and of the LFSR are so small that they cannot be used for correlation attacks.

In addition, a filtering based on a function f(x) of degree 7 is not vulnerable to algebraic attacks as the algebraic degrees of the output bits when expressed as a function of LFSR-bit are large in general, and varying in time so this defeats algebraic attacks. As well, the cost of time/memory/data tradeoff attacks on stream ciphers is $O(2^{L/2})$, where L is the number of inner states of the stream cipher. To comply with the margins set by this attack, L=16 has been chosen. The sampling resistance of f(x) is reasonable because this function does not become linear in the remaining variables by fixing less than half of its 16 variables.

According to the above analysis, sequences produced by the proposed PRNG pass most statistical analyses, have short hardware requirements and are resistant to known attacks, what confirms its validity and the hypothesis that the proposed nonlinear filtering and decimation solve the linearity problem of LFSR-based generators. We also showed that the proposal meets all the requirements of the EPC Gen2 standard, what leads us to defining a new mutual authentication scheme for EPC Gen2 in the next section.

## 6. Proposed Authentication Scheme

One of the most interesting and necessary studies in RFID is the development of strong authentication schemes for RFID systems with resource-constrained tags. A new mechanism to provide authenticity for low-cost RFID systems fulfilling EPC Gen2 is proposed below. This is a difficult task due to the relative ease with which an adversary can record or participate in a conversation between tag and reader.

The proposed method can be used both by reader and tag to mutually authenticate each other and to establish a shared session secret key. It assumes that the reader is linked through a secure communication channel to a back-end server with a database where each tag t is related to EPCdata formed by a 16-bit secret identification number $ID_{t,i}$ and a 16-bit shared secret key $SSK_{t,i}$ for each phase i=1,2, ... It is also assumed that both reader and tag are able to use the proposed PRNG. Both the identification number $ID_{t,i}$ and the shared secret key $SSK_{t,i}$ are updated through the PRNG, and nonces of length $n<2^{16}$ are used.

Below the steps of the algorithm proposed for both mutual authentication and session key establishment are described, as shown in Figure 5.

**Algorithm:**

1. The reader sends a random Query message of length 16 to the tag.
2. The tag t feeds the PRNG with ($ID_{t,i}$ XOR Query) and with ($SSK_{t,i}$ XOR Query) to produce two (16+n)-bit keystream sequences whose last n bits are XORED to be sent, together with a random 16-bit NONCE1, to the reader.
3. The reader sends the data received in Step 2 to the back-end server, which compares them with all outputs corresponding to the stored pairs ($ID_{t,j}$, $SSK_{t,j}$).
4. If the server finds no collision, it identifies it as a possible fraud attempt. Otherwise, if it finds only one collision for ($ID_{t,i}$, $SSK_{t,i}$), then it sends to the reader both the session key K= $ID_{t,i}$ XOR $SSK_{t,i}$, and the result NONCE3 of the XOR between the last n bits of the two (16+n)-bit keystream sequences produced by the PRNG on ($ID_{t,i}$ XOR NONCE1) and on ($SSK_{t,i}$ XOR NONCE1), and updates t's data to ($ID_{t,i+1}$, $SSK_{t,i+1}$). Else, if the server finds more than one collision (although the probability of this case is negligible), the server informs the reader about the failure so that the process is restarted from the first step.

5. The reader sends to the tag the received sequence NONCE3.

According to the above scheme the following properties are fulfilled:

- In the last step the tag checks whether the received data coincides with the sequence produced by itself on the correct data, and if this verification is successful it updates its EPCdata to $(ID_{t,i+1}, SSK_{t,i+1})$
- After the execution of the five steps in the above scheme, both the reader and the tag can use the same secret session key $K = ID_{t,i}$ XOR $SSK_{t,i}$.
- The information that an eavesdropper of the insecure channel between reader and tag obtain from repeated listening is useless.

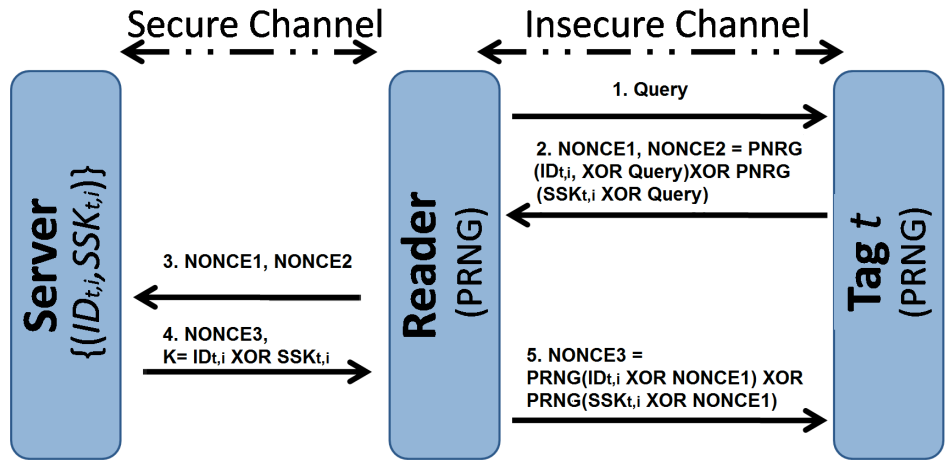

Fig. 5 Mutual Authentication Scheme

The established session key K will be then used both by tag and reader to initialize the stream cipher generator PRNG to obtain the same key stream Z to encrypt all messages exchanged between them during that session. During that session they could also use such a shared key for fast challenge-response authentication based on symmetric cryptography.

In ubiquitous environments we can assume that not many problems of connectivity exist, so for simplicity and practical security in our proposal we have assumed the existence of continuous and secure connectivity between readers and back-end server, and desynchronization is not considered. Note that in step 3 in which the server must compare the output of the PRNG over all EPCdata of the tags it manages, the CRC can be very useful for finding collisions in its database. Also, it is remarkable that the scheme does not provide any information useful for potential eavesdroppers of messages exchanged between tag and reader thanks to the fact that the PRNG is a non-linear function. On the other hand, without knowledge of the corresponding EPCdata of the tag, it is very difficult to build a value that the server can recognize as valid. Therefore the proposed protocol actually provides tag authentication. Regarding tag privacy protection and tracking attack prevention, the proposed protocol protects both qualities because the response of the tag in step 2 is different and unpredictable in each authentication request due to the update of its EPCdata. Note also that the update of its secret identification number and its shared secret key involves forward security feature and resistance against replay attacks. In addition, neither the tag nor the server provides ever the ID to the reader; therefore there is no possibility that a legitimate and malicious reader can perform impersonation attacks against any tag.

Man In The Middle (MITM) attacks are impossible to cope with the proposal because they require that the attacker can make independent connections with the reader and the tag in order to relay messages between them to make them believe that they are talking directly to each other when, in fact, the entire conversation is controlled by the attacker. In order to be successful in MITM attacks, the attacker should be able to intercept all messages between tag and reader and inject new ones to impersonate the tag in front of the reader and the reader in front of the tag, but this is not possible in our proposal as either the server or the tag detect the attack due to the ignorance of the attacker about the tag's data. Thus, if the server detects an attack, it informs the reader that the message received in step 2 does not produce any collision. If with a negligible probability the server finds some random collision after a MITM attack, then the tag detects the attack when from the message received in step 5 the data does not correspond to the correct data. Thus, we can conclude that the proposed mutual authentication scheme is immune front MITM attacks and provides authentication of reader.

## 7. Conclusions

In this work, two new cryptographic primitives that meet the EPC Gen2 standard of RFID technology have been proposed. On the one hand, a new PRNG whose design is based on a nonlinear filtering of a LFSR has been described and analyzed. The obtained results confirm the good randomness and non-linearity properties of the output sequences. On the other hand, we have proposed a lightweight mutual authentication scheme that is based on the proposed generator. Both systems meet all practical requirements of low-cost RFID, and security properties such as confidentiality, mutual authentication and anonymity of tags. Thus, both schemes are immune against known attacks on PRNGs and authentication schemes. In conclusion, the proposals might be considered a reference solution for security problems of EPC Gen2 RFID. Several topics deserve further study. For instance, formal security proofs for the protocols could be developed. Also, the authentication scheme could be adapted so that it can be used when the channel between the back-end server and the reader is insecure.

**Acknowledgments**

Research supported by the Ministerio de Ciencia e Innovación and the European FEDER Fund under TIN2008-02236/TSI and BES-2009-016774, and by the Agencia Canaria de Investigación, Innovación y Sociedad de la Información under PI2007/005 and BOC 60.

*Corresponding author: Pino Caballero-Gil, Ph.D.

University of La Laguna, Spain. Email: pcaballe@ull.es